\documentclass{ametsocV6}
\usepackage[number-unit-product=~]{siunitx}
\usepackage{xcolor}

\newcommand{\pder}[2]{\frac{\partial#1}{\partial#2}}
\newcommand{\B}{_\mathrm{B}}
\newcommand{\I}{_\mathrm{I}}
\renewcommand{\vec}{\mathbf}
\newcommand{\Izero}{_{\mathrm{I}0}}

\newcommand{\Itwo}{_{\mathrm{I}2}}
\newcommand{\Bzero}{_{\mathrm{B}0}}
\newcommand{\Bone}{_{\mathrm{B}1}}
\newcommand{\Btwo}{_{\mathrm{B}2}}

\defcitealias{peterson_rapid_2022}{PC22}

\title{Coupling Between Abyssal Boundary Layers and the Interior Ocean \\ in the Absence of Along-Slope Variations}
\authors{Henry G. Peterson\correspondingauthor{hgpeterson@caltech.edu} and J\"orn Callies}
\affiliation{California Institute of Technology, Pasadena, California}

\abstract{To close the overturning circulation, dense bottom water must upwell via turbulent mixing.
Recent studies have identified thin bottom boundary layers (BLs) as locations of intense upwelling, yet it remains unclear how they interact with and shape the large-scale circulation of the abyssal ocean.
The current understanding of this BL--interior coupling is shaped by 1D theory, suggesting that variations in locally produced BL transport generate exchange with the interior and thus a global circulation.
Until now, however, this picture has been based on a 1D theory that fails to capture the local evolution in even highly idealized 2D geometries.
The present work applies BL theory to revised 1D dynamics, which more naturally generalizes to two and three dimensions.
The BL is assumed to be in quasi-equilibrium between the upwelling of dense water and the convergence of downward buoyancy fluxes.
The BL transport, for which explicit formulae are presented, exerts an influence on the interior by modifying the bottom boundary condition.
In 1D, this BL transport is independent of the interior evolution, but in 2D the BL and interior are fully coupled.
Once interior variables and the bottom slope are allowed to vary in the horizontal, the resulting convergences and divergences in the BL transport exchange mass with the interior.
This framework allows for the analysis of previously inaccessible problems such as the BL--interior coupling in the presence of an exponential interior stratification, laying the foundation for developing a full theory for the abyssal circulation.}

\begin{document}

\maketitle

\section{Introduction}

Thin boundary layers (BLs) at the ocean's bottom have recently come into focus as the primary locations in which small-scale turbulence lightens bottom waters, thus playing a crucial role in closing the overturning circulation of the abyss \citep{ferrari_turning_2016,de_lavergne_consumption_2016}. 
The connection between these BLs and the large-scale abyssal circulation, however, remains to be fully explained.
The cornerstone of our present understanding of the mixing-generated abyssal circulation is a 1D model of a stratified, rotating fluid overlying a sloping, insulated seafloor \citep[e.g.,][]{phillips_flows_1970,wunsch_oceanic_1970,thorpe_current_1987,garrett_boundary_1993}.
This 1D theory helped bring bottom BLs into center stage, predicting that the local response to bottom-intensified mixing is characterized by diabatic upslope flow in the thin BL compensated in part by diabatic downslope flow spread across the interior \citep{garrett_role_1990,ferrari_turning_2016,de_lavergne_consumption_2016,mcdougall_abyssal_2017,callies_restratification_2018}. 
Our description of large-scale abyssal dynamics is shaped by this local theory: the natural conclusion is that variations in these locally produced flows generate exchange with the interior and producing a global circulation \citep[e.g.,][]{phillips_experiment_1986,mcdougall_dianeutral_1989,garrett_marginal_1991,dell_diffusive_2015,holmes_ridges_2018}.
This picture fails to consider the potential feedback of the circulation produced in the interior back onto the BL, however, suggesting that this framework is incomplete.

In addition to this lack of two-way coupling, progress has also been hampered by the canonical 1D theory failing to reproduce the local evolution in simple 2D geometries.
The canonical 1D model predicts slow diffusion of the interior along-slope flow \citep{maccready_buoyant_1991}, whereas simulations of bottom-intensified mixing over an idealized 2D mid-ocean ridge display rapid spin up of the interior \citep{ruan_mixing-driven_2020}.
In \citet[][hereafter \citetalias{peterson_rapid_2022}]{peterson_rapid_2022}, we remedied this shortcoming by including the physics of a secondary circulation and barotropic pressure gradient.
The key is to constrain the vertically integrated cross-slope transport to force upwelling flow in the BL to return in the interior.
This downwelling flow is then turned in the along-slope direction by the Coriolis acceleration and balanced by a barotropic pressure gradient, leading to rapid adjustment in the interior as seen in 2D.
With this more faithful 1D model, we have a reliable foundation to describe the role of abyssal BLs in the large-scale circulation.

\citet{callies_dynamics_2018} and \citet{drake_abyssal_2020} connected BL dynamics to the horizontal circulation in a 3D planetary-geostrophic (PG) model with idealized bathymetry and Rayleigh friction.
\citet{callies_dynamics_2018} found that, for vertically constant interior stratification and on moderate slopes, local 1D theory accurately emulates the 3D model's dynamics.
On the sloping sidewalls of the idealized bathymetry, upslope transport in thin bottom BLs is compensated by downwelling aloft.
At the base of the slopes, however, 1D theory breaks down in favor of a basin-scale circulation that feeds the BLs on slopes.
An integral of the local upslope 1D BL transport along the perimeter of the basin provides an accurate estimate of the overturning.
These ideas fail, however, once the interior stratification is far from constant, because 1D theory can only consider perturbations to a constant background stratification \citep{drake_abyssal_2020}.
This is a severe limitation, given the real ocean's near-exponential stratification \citep[e.g.,][]{munk_abyssal_1966}.
For a more realistic stratification, downwelling in the interior is weakened and BL upwelling dominates, though the vertical extent and structure of the net transport remains to be explained.
In this work, we provide a framework for concretely understanding this interplay between the BL and interior.

\begin{figure}
    \centering
    \includegraphics[width=19pc]{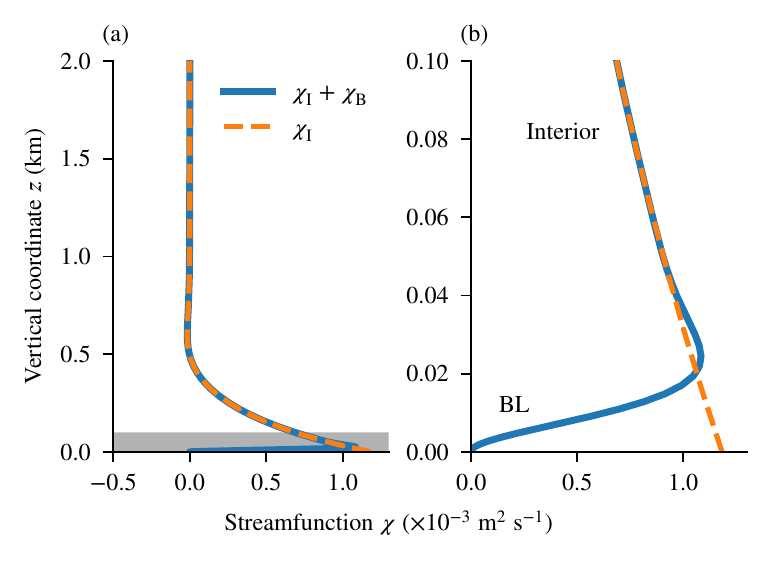}
    \caption{Illustration of the BL correction to interior solution.
    Shown is a typical streamfunction~$\chi$, defined such that~$\partial_z \chi = u^x$ where~$u^x$ is the cross-slope flow, after three years of mixing-generated abyssal spin up at a slope Burger number~$\varrho = \num{e-3}$ (see section~\ref{s:1d}).
    The solution is depicted over~(a)~the entire~\SI{2}{\kilo\meter} domain as well as~(b)~a zoom-in to the bottom~\SI{100}{\meter}, shown in (a) in gray shading.
    The interior solution~$\chi\I$ varies slowly compared with the scale of the BL, and the BL correction~$\chi\B$ ensures that boundary conditions are satisfied.
    }
    \label{fig:BL_correction}
\end{figure}

Below, we derive self-contained equations for interior 1D and 2D PG dynamics on an $f$-plane with effective boundary conditions that capture the effects of BLs.
We accomplish this using BL theory, splitting variables into their interior and BL contributions \citep[e.g.,][Fig.~\ref{fig:BL_correction}]{bender_advanced_1999,chang_advanced_2007}.
This explicitly separates the interior and BL dynamics and allows for deep physical insight into their coupling. 
Famously, \citeauthor{stommel_westward_1948}'s (\citeyear{stommel_westward_1948}) gyre theory can be solved with BL methods \citep{veronis_wind-driven_1966}, although the coupling there is one-way: the interior solution can be calculated in isolation, and the western BL is a passive element of the theory. 
We find that this is different for bottom BLs on slopes. 
Their structure is shaped by the interior solution, but the buoyancy and mass fluxes carried in the BL feed back on the interior solution in the form of boundary conditions. 

A central result of this paper is an explicit expression for the cross-slope BL transport (per unit along-slope distance) in terms of interior variables and flow parameters.
In 1D, the BL transport takes the form $\kappa \cot\theta \, \mu \varrho / (1 + \mu \varrho)$, where $\mu = \nu/\kappa$ is the turbulent Prandtl number with~$\nu$ being the turbulent viscosity and~$\kappa$ the turbulent diffusivity, and~$\varrho = N^2 \tan^2\theta / f^2$ is the slope Burger number with~$N$ being the background interior buoyancy frequency, $f$ the inertial frequency, and~$\theta$ the bottom slope angle.
All variables are evaluated at the bottom (or, more generally, just above the BL).
In the canonical 1D framework, a steady-state balance between cross-slope upwelling of dense water and turbulent mixing requires that the \textit{total} transport tends towards~$\kappa_\infty \cot\theta$, where $\kappa_\infty$ is the far-field turbulent diffusivity \citep{thorpe_current_1987,garrett_boundary_1993}.
Our revised result instead applies to the bottom BL transport and is valid throughout transient evolution, provided that the BL has adjusted to a quasi-steady state. 
Unlike the canonical result, this expression smoothly approaches zero as~$\theta \to 0$, more harmoniously connecting the model over a slope with conventional flat-bottom Ekman theory \citep[e.g.,][]{pedlosky_geophysical_1979}.
The expression has the same form in 2D, but there the slope Burger number is a function of interior cross-isobath buoyancy gradients as well as the local topographic slope.
Thus, in 2D, variations in interior buoyancy gradients and the topographic slope cause convergence in the BL transport, generating exchange with the interior.
A similar process occurs in 3D with the added physics of along-isobath variations and a modified interior balance, but we leave the details of 3D dynamics to future work.

In section~\ref{s:1d}, we begin by reviewing the transport-constrained 1D model from \citetalias{peterson_rapid_2022}, followed by a derivation of the 1D BL theory.
We derive the 2D BL theory in section~\ref{s:2d}, applying the framework to simulations of mixing-generated spin up under a vertically varying background stratification.
In section~\ref{s:asymptotics}, we re-derive the 1D and 2D BL equations in a more rigorous fashion, quantifying the accuracy of our claims in the previous sections and uncovering some subtleties in the dynamics.
Finally, we provide discussion and conclusions in sections~\ref{s:discussion} and~\ref{s:conclusions}, respectively.

\section{One-dimensional boundary layer theory}\label{s:1d}

In this section, we apply BL theory to the revised 1D model from \citetalias{peterson_rapid_2022} and present results from numerical integrations of both the full and BL equations. Here and throughout the paper, we employ PG scaling, thus focusing our attention on the slow and large-scale response to mixing.
The PG flow should be interpreted as the residual flow after a thickness-weighted average over transients due to turbulence, waves, and baroclinic eddies, with the effect of these transients included as parameterized Eliassen--Palm and diapycnal fluxes \citep{young_exact_2012}.

\subsection{Transport-constrained one-dimensional dynamics}

We first consider 1D PG dynamics along a uniform slope at an angle~$\theta$ above the horizontal.
The 1D model is typically derived by writing the Boussinesq equations in a rotated coordinate system aligned with the slope \citep[e.g.,][]{garrett_boundary_1993}.
We slightly deviate from this approach by keeping the vertical coordinate aligned with gravity, which is a more natural choice if the horizontal components of the turbulent momentum and buoyancy fluxes are neglected, but it yields equivalent dynamics \citepalias{peterson_rapid_2022}.\footnote{In the limit $\theta \ll 1$, the gravity-aligned coordinate system employed here and the previously used fully rotated coordinate system yield the same equations.}
Specifically, we write the 1D model in~$(\xi, \eta, \zeta)$ coordinates defined by
\begin{equation}
    \xi = x, \quad \eta = y, \quad \zeta = z - x \tan \theta,
\end{equation}
where~$(x, y, z)$ defines the usual Cartesian coordinate system with~$z$ aligned with gravity.
These coordinates are analogous to terrain-following coordinates (used below) in 1D with~$\zeta = 0$ at the bottom.
Neglecting all variations in~$\xi$ and~$\eta$, except for the barotropic pressure gradient~$\partial_x P$ (equivalently, $\partial_\xi P$, since $P$ is independent of~$z$), and constraining the vertically integrated cross-slope transport to~$U^\xi$ (typically to zero), the PG equations become
\begin{align}
    -fu^\eta &= -\pder{P}{x} + b'\tan\theta + \pder{}{\zeta}\left(\nu \pder{u^\xi}{\zeta} \right)\label{eq:1d-xi},\\
    fu^\xi &= \pder{}{\zeta}\left(\nu \pder{u^\eta}{\zeta}\right)\label{eq:1d-eta},\\
    \pder{b'}{t} + u^\xi N^2\tan\theta &= \pder{}{\zeta}\left[\kappa\left(N^2 + \pder{b'}{\zeta}\right)\right]\label{eq:1d-b}, \\
    \int_0^\infty u^\xi \; d\zeta &= U^\xi\label{eq:1d-U}.
\end{align}
Here, $u^\xi$ is the cross-slope velocity\footnote{Due to our non-orthogonal coordinate system, $u^\xi$ is technically the $x$-projection of the cross-slope velocity as it would be defined in a fully rotated coordinate system \citepalias[][appendix A]{peterson_rapid_2022}. For simplicity, we refer to it as the ``cross-slope velocity'' throughout.} and~$u^\eta$~is the along-slope velocity.
We have split the total buoyancy~$b$ into a constant background stratification and a perturbation so that~$b = N^2 z + b'$.
The fluid satisfies no-slip and insulating boundary conditions at the bottom: $u^\xi = 0$, $u^\eta = 0$, and~$\partial_\zeta b = N^2 + \partial_\zeta b' = 0$ at~$\zeta = 0$.
In the far field, we impose decay conditions on the shear and anomalous buoyancy flux: $\partial_\zeta u^\xi \to 0$, $\partial_\zeta u^\eta \to 0$, and~$\partial_\zeta b' \to 0$ as~$\zeta \to \infty$.
The extra degree of freedom supplied by~$\partial_x P$ allows the transport constraint~\eqref{eq:1d-U} to be satisfied at all times.
Physically, this constraint forces cross-slope upwelling in the BL to return in the interior, where it is then turned into the along-slope direction by the Coriolis force.
In the PG framework, this process is instantaneous, and the far-field along-slope flow satisfies the balance: $-fu^\eta = -\partial_x P$.
This leads to rapid spin up of the along-slope flow throughout the water column, as seen in simulations of 2D spin up \citep[][\citetalias{peterson_rapid_2022}]{ruan_mixing-driven_2020}.

We employ a simple down-gradient closure for the turbulent momentum and buoyancy fluxes generated by, e.g., breaking internal waves but allow for variations in the mixing coefficients~$\nu$ and~$\kappa$.
We assume these variations to occur on a scale larger than the BL thickness.
In our examples below, $\nu$ and~$\kappa$ are bottom-enhanced in abyssal mixing layers a few hundred meters thick, inspired by typical observations over rough mid-ocean ridges.
Our main results, however, generalize to the case in which~$\nu$ and~$\kappa$ vary rapidly within the BL, for example going to zero in a log-layer.

As in \citetalias{peterson_rapid_2022}, we cast equations~\eqref{eq:1d-xi} to~\eqref{eq:1d-U} into an inversion equation for the flow, written in terms of a streamfunction~$\chi(\zeta)$ defined such that~$u^\xi = \partial_\zeta \chi$, and an evolution equation for the buoyancy perturbation:
\begin{align}
    \pder{^2}{\zeta^2}\left(\nu\pder{^2\chi}{\zeta^2}\right) + \frac{f^2}{\nu}(\chi - U^\xi) &= -\pder{b'}{\zeta}\tan\theta,\label{eq:1d-inversion}\\
    \pder{b'}{t} + \pder{\chi}{\zeta} N^2\tan\theta &= \pder{}{\zeta} \left[\kappa\left(N^2 + \pder{b'}{\zeta}\right)\right].\label{eq:1d-evolution}
\end{align} 
The boundary conditions are that~$\chi = 0$ and~$\partial_\zeta \chi = 0$ at~$\zeta = 0$ and~$\chi \to U^\xi$ as~$\zeta \to \infty$. 
If desired, one may infer the along-slope flow from~$\chi$ by integrating
\begin{equation}\label{eq:1d-ueta-from-chi}
    \pder{u^\eta}{\zeta} = \frac{f}{\nu}(\chi - U^\xi)
\end{equation}
from the bottom up, using~$u^\eta = 0$ at~$\zeta = 0$.
Equations~\eqref{eq:1d-inversion} and~\eqref{eq:1d-evolution} fully describe the 1D PG system and can readily be solved numerically. 
But insight into the BL--interior coupling is more easily gained using BL theory.

\subsection{Boundary layer theory}

\begin{figure*}
    \centering
    \includegraphics[width=33pc]{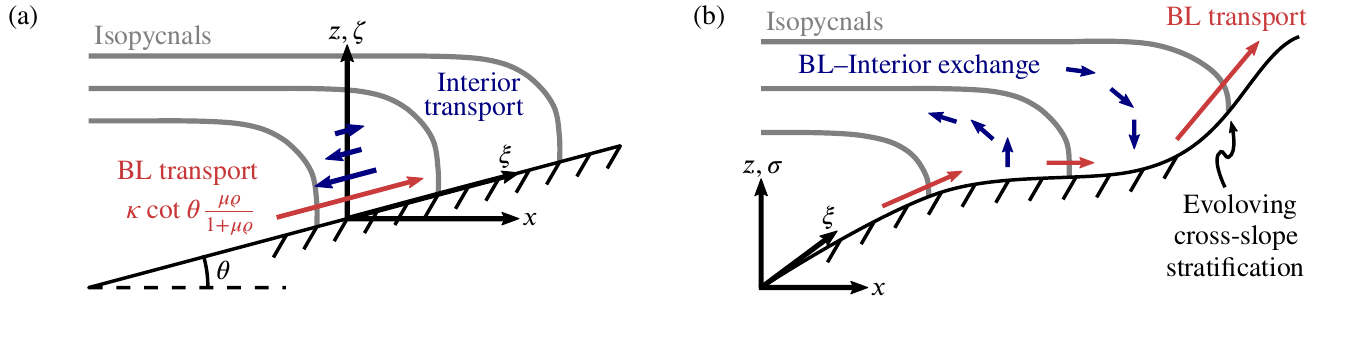}
    \caption{Sketch of BL theory framework for (a)~1D dynamics over a uniform slope and (b)~2D dynamics over more complicated topography.}
    \label{fig:BL_sketch}
\end{figure*}

Under steady conditions, equations~\eqref{eq:1d-inversion} and~\eqref{eq:1d-evolution} can be combined to form a single fourth-order ordinary differential equation for~$\chi$.
The fourth- and zeroth-order terms in that equation balance if~$\chi$ varies on a scale~$q^{-1}$ defined by
\begin{equation}\label{eq:q}
    (\delta q)^4 = 1 + \mu \varrho,
\end{equation}
where~$\delta = \sqrt{2\nu / f}$ is the familiar flat-bottom Ekman layer thickness, and the mixing coefficients are evaluated at~$\zeta = 0$.
This defines the BL scale of a rotating fluid adjacent to a sloping bottom \citep[e.g.,][]{garrett_boundary_1993}.
For typical abyssal parameters, $q^{-1} \sim \SI{10}{\meter}$ \citep{callies_restratification_2018}.
This thinness of the BL compared to the scale of variations in the interior ocean is what allows us to apply BL theory.

We begin by splitting solutions into interior contributions~$\chi\I$ and~$b'\I$, which vary slowly in~$\zeta$, and BL corrections~$\chi\B$ and~$b'\B$, which ensure boundary conditions are satisfied and have appreciable magnitude in the thin BL only.
A similar approach was taken in \citet{callies_restratification_2018} with the canonical 1D model, but the analysis presented here is time-dependent and extensible to higher dimensions (section~\ref{s:2d}).
If the mixing coefficients~$\nu$ and~$\kappa$ vary on a scale much larger than~$q^{-1}$, the fourth-order term in~\eqref{eq:1d-inversion} can be neglected in the interior:
\begin{equation}\label{eq:1d-inversion-interior}
    \frac{f^2}{\nu}\chi\I = -\pder{b'\I}{\zeta}\tan\theta,
\end{equation}
assuming~$U^\xi = 0$ (see appendix~A for the $U^\xi \neq 0$ case).
Substituted back into the buoyancy equation~\eqref{eq:1d-evolution}, this reduces the interior dynamics to a modified diffusion equation:
\begin{equation}
    \pder{b'\I}{t} = \pder{}{\zeta}\left(\kappa\left[N^2 + (1 + \mu \varrho)\pder{b'\I}{\zeta}\right]\right).
    \label{eq:1d-evolution-interior}
\end{equation}
This is a result familiar from \citet{gill_homogeneous_1981}, \citet{garrett_dynamical_1981}, and \citet{garrett_spindown_1982}: advection of the background stratification by the secondary circulation becomes a horizontal diffusion term, with diffusivity~$\nu N^2 / f^2$. 
The form here is the result of the sloping boundary: the vertical coordinate depends on the slope-parallel distance multiplied by~$\tan \theta$, which explains the factor~$\tan^2\theta$ in the additional diffusion term.

This interior evolution must be complemented by a representation of the bottom BL that supplies an effective boundary condition for the interior equation.
The key assumption here is that the BL scale~$q^{-1}$ is thin compared to interior variations. 
This thinness of the BL also implies that it is quasi-steady on the time scales of the interior evolution. 
The BL correction thus satisfies the steady buoyancy equation
\begin{equation} \label{eq:1d-bl-buoyancy}
    \pder{\chi\B}{\zeta} N^2 \tan \theta = \pder{}{\zeta} \left( \kappa \pder{b'\B}{\zeta} \right).
\end{equation}
Since all BL variables decay into the interior, i.e., as~$\zeta \to \infty$, this balance can be integrated to 
\begin{equation}
    \chi\B N^2 \tan\theta = \kappa \pder{b'\B}{\zeta}.
\end{equation}
This relation is all that is needed to derive a boundary condition on the interior solution.
At~$\zeta = 0$, $\chi\I + \chi\B = 0$, such that the full $\chi = 0$ boundary condition is satisfied. 
So, using \eqref{eq:1d-inversion-interior},
\begin{equation}
    \pder{b'\B}{\zeta} = -\frac{N^2\tan\theta}{\kappa} \chi\I = \mu \varrho \pder{b'\I}{\zeta} \quad \text{at} \quad \zeta = 0.
\end{equation}
The insulating boundary condition then becomes
\begin{equation}
    0 = N^2 + \pder{b'\I}{\zeta} + \pder{b'\B}{\zeta} = N^2 + (1 + \mu \varrho) \pder{b'\I}{\zeta} \quad \text{at} \quad \zeta = 0.
    \label{eq:1d-interior-bc}
\end{equation}
The BL correction thus contributes an additional term~$\mu \varrho \partial_\zeta b'\I$ to the boundary condition for the interior buoyancy evolution~\eqref{eq:1d-evolution-interior}. 
The added term represents physics akin to an Ekman buoyancy flux \citep[e.g.,][]{marshall_fluid_1992,thomas_intensification_2005}: the BL transport~$\chi\I$ acts on the cross-slope buoyancy gradient~$N^2 \tan\theta$ and produces a buoyancy sink for the interior.
This boundary condition on the interior problem implies a stratification at the top of the BL that is reduced from the background by a factor~$\mu \varrho /(1 + \mu \varrho)$ and a BL transport, from combining~\eqref{eq:1d-interior-bc} and~\eqref{eq:1d-inversion-interior},
\begin{equation}\label{eq:1d-bl-transport}
    \chi\I = \kappa \cot\theta \frac{\mu \varrho}{1 + \mu \varrho} \quad \text{at} \quad \zeta = 0,
\end{equation}
as claimed in the introduction (Fig.~\ref{fig:BL_sketch}a).
We note that the transport-constrained system, unlike the canonical one, has no steady state in a semi-infinite
domain, yet previous work on the BL--interior interaction has often begun with the canonical result that the steady transport is~$U^\xi = \kappa_\infty \cot \theta$ \citep[e.g.,][]{woods_boundary-driven_1991,callies_dynamics_2018,drake_abyssal_2020}.
The revised expression in~\eqref{eq:1d-bl-transport} instead applies to the transport confined to the BL and more sensibly leaves the net transport (and steady-state dynamics) to be controlled by the large-scale context.

If desired, the BL correction can easily be determined from
\begin{equation}\label{eq:1d-bl-equation}
    \pder{^4\chi\B}{\zeta^4} + 4q^4\chi\B = 0,
\end{equation}
with~$\chi\B = -\chi\I$ and~$\partial_\zeta \chi\B = 0$ at~$\zeta = 0$ (neglecting the much smaller interior contribution to~$\partial_\zeta \chi$ at the bottom) and~$\chi\B \to 0$ as~$\zeta \to \infty$.
This has a similar form as the steady canonical 1D problem with constant mixing coefficients \citep[e.g.,][]{garrett_boundary_1993}, but the boundary conditions and right-hand side are different because the transport constraint is imposed and the interior solution has been subtracted out.
The general solution takes the form of the familiar Ekman spiral:
\begin{equation}\label{eq:1d-bl-sol}
    \chi\B = -\chi\I e^{-q \zeta} ( \cos q\zeta + \sin q\zeta ),
\end{equation}
where~$\chi\I$ is evaluated at~$\zeta = 0$ as in~\eqref{eq:1d-bl-transport}.

This analytical expression for the BL correction also allows us to directly diagnose how the far-field along-slope flow is influenced by the BL.
From~\eqref{eq:1d-ueta-from-chi} and~\eqref{eq:1d-inversion-interior}, the interior along-slope shear follows thermal wind balance,
\begin{equation}
    \pder{u^\eta\I}{\zeta} = -\frac{1}{f} \pder{b'\I}{\zeta} \tan\theta,
\end{equation}
which implies, upon integration in the vertical,
\begin{equation}\label{eq:1d-ueta-interior}
    u^\eta\I(\zeta) = u^\eta\I(0) - \frac{1}{f} \left[b'\I(\zeta) - b'\I(0) \right] \tan\theta.
\end{equation}
The integration constant~$u^\eta\I(0)$, the flow at the upper edge of the BL, can be determined from the BL solution~\eqref{eq:1d-bl-sol} and~\eqref{eq:1d-ueta-from-chi}: $u^\eta\I(0) = -u^\eta\B(0) = -f \chi\I(0) / q \nu(0)$.
This BL contribution to the interior along-slope flow has the same form as the steady-state canonical result with constant mixing coefficients \citep{thorpe_current_1987,garrett_boundary_1993}, but here it is rapidly spun up and accompanied by an additional interior thermal-wind component.
We will see in section~\ref{s:asymptotics} that this BL contribution is typically of higher asymptotic order than the thermal-wind contribution.

It should be noted that the key results \eqref{eq:1d-interior-bc} and~\eqref{eq:1d-bl-transport} also apply if there are variations in the mixing coefficients within the thin BL, as may be expected as the turbulence becomes suppressed very close to the bottom. The physics that lead to~\eqref{eq:1d-interior-bc} and~\eqref{eq:1d-bl-transport} are that the diffusive buoyancy flux into the BL is balanced by cross-slope advection within the BL and that the interior obeys~\eqref{eq:1d-inversion-interior}. While the BL corrections are more complicated if $\nu$ and $\kappa$ are not approximately constant across the BL, for example including a log-layer if the mixing coefficients go to zero near the bottom, the effective boundary condition for the interior is the same.

In summary, BL theory has enabled us to elucidate the connection between the BL and interior in 1D.
The BL transport quickly adjusts to~\eqref{eq:1d-bl-transport}, regardless of the interior dynamics.
This transport allows the BL to communicate with the interior by moving dense water up the slope, providing a buoyancy sink and modifying the interior bottom boundary condition~\eqref{eq:1d-interior-bc} (Fig~\ref{fig:BL_sketch}a).
In 1D, the BL is thus independent of the evolution of the interior, yet the cross-slope advection by the BL transport affects the interior dynamics.
As we will see in the next section, the BL--interior coupling in 2D are even richer, with the interior being able to feed back onto the BL.
But first, we present some illustrative 1D examples.

\subsection{Examples}

\begin{table}[b]
    \centering
    \begin{tabular}{l c c}
        \hline\hline
        Inertial frequency &~$f$ & \SI{-5.5e-5}{\per\second}\\
        Far-field buoyancy frequency &~$N$ & \SI{e-3}{\per\second}\\
        Far-field diffusivity &~$\kappa_0$ & \SI{6e-5}{\meter\squared\per\second}\\
        Bottom-enhancement of diffusivity &~$\kappa_1$ & \SI{2e-3}{\meter\squared\per\second}\\
        Decay scale of diffusivity &~$h$ & 200 m\\
        Prandtl number &~$\mu$ & 1\\
        \hline
    \end{tabular}
    \caption{Parameters used in simulations of spin up, adapted from \citet{callies_dynamics_2018} and roughly corresponding to the mid-Atlantic ridge flank in the Brazil Basin.
    }
    \label{tab:params}
\end{table}

\begin{figure*}
    \centering
    \includegraphics[width=27pc]{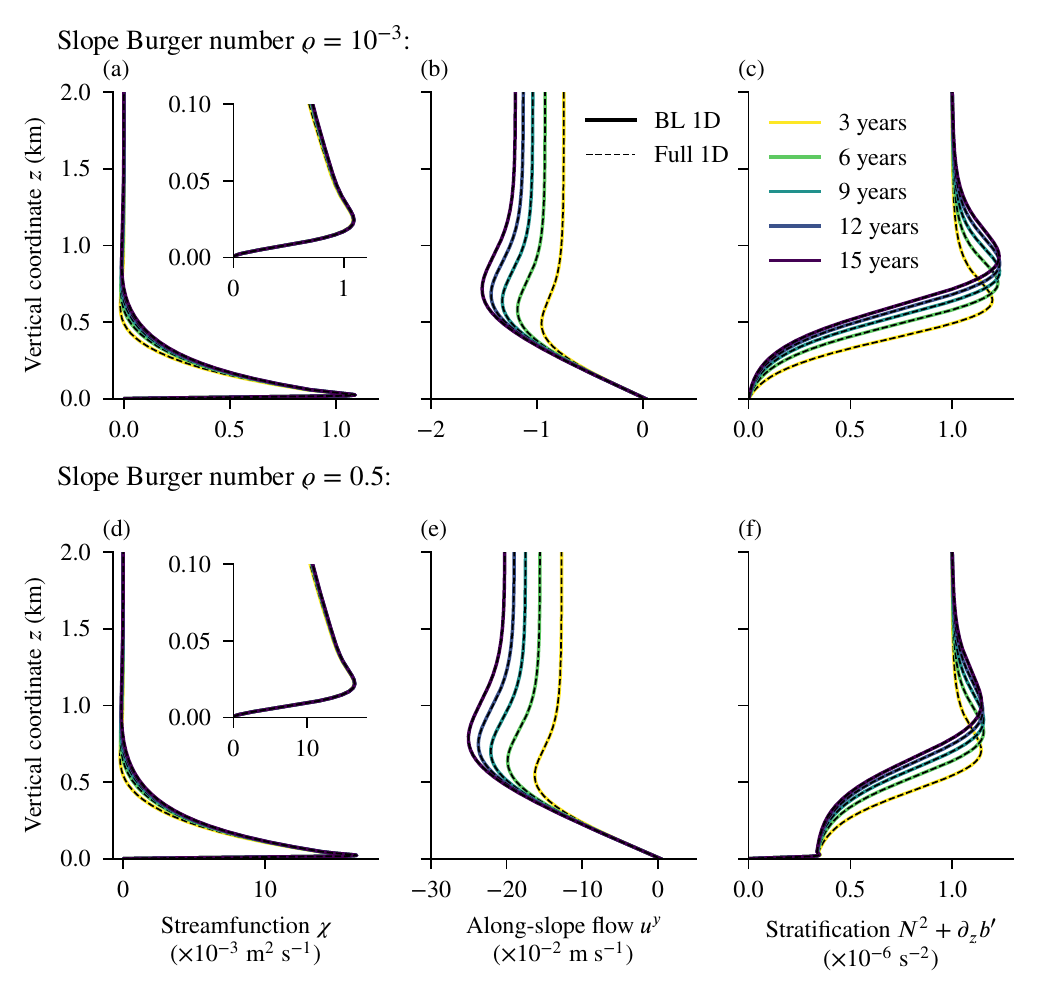}
    \caption{Comparison of the 1D BL solution with full 1D PG spin up over two different slope angles.
    Shown are the (a), (d)~streamfunction~$\chi$, (b), (e)~along-slope flow~$u^y = u^\eta$, and (c), (f)~stratification~$N^2 + \partial_z b'$ as functions of~$z = \zeta$ for separate simulations in which the slope Burger number is (a--c)~$\varrho = \num{e-3}$, corresponding to a bottom slope of~$\theta \approx \SI{1.7e-3}{\radian}$, and (d--f)~$\varrho = 0.5$ so that~$\theta \approx \SI{3.9e-2}{\radian}$.
    The insets of (a) and (d) show the streamfunction~$\chi$ in the bottom~\SI{100}{\meter}, showcasing the accuracy of the BL correction.
    The 1D BL theory matches the 1D dynamics perfectly.
    }
    \label{fig:slope_profiles}
\end{figure*}

The following experiments depict 1D PG spin up with and without BL theory.
The simulations start in a state of rest: isopycnals are flat ($b' = 0$), and the flow is zero ($\chi = 0$).
The turbulent mixing then generates a buoyancy perturbation, bending isopycnals into the slope and spinning up a circulation.
The transport constraint ensures that BL transport is exactly returned in the interior, and without a source of dense bottom water, the initial stratification is mixed away with time.

To numerically solve the 1D PG equations, we use second-order finite differences as in \citetalias{peterson_rapid_2022}.
The model can either solve for the full flow and density profiles using equations~\eqref{eq:1d-inversion} and~\eqref{eq:1d-evolution} or evolve the interior variables of the BL theory with equation~\eqref{eq:1d-evolution-interior}.
Model parameters are adapted from \citet{callies_restratification_2018} and roughly match those of the Brazil Basin (Table~\ref{tab:params}).
Mixing is represented by a bottom-intensified profile of turbulent diffusivity, 
\begin{equation}
    \kappa = \kappa_0 + \kappa_1 e^{-\zeta/h},
\end{equation}
with parameters obtained from a fit to Brazil Basin observations \citep[][Table~\ref{tab:params}]{callies_restratification_2018}.
When solving the full 1D PG equations, grid spacing follows Chebyshev nodes with resolution on the order of 0.1~m at~$\zeta = 0$ to comfortably resolve the boundary layers.
The BL simulations need not resolve the thin bottom BL, and we therefore use a uniform grid spacing of~\SI{8}{\meter} for these.
The domain height of~\SI{2}{\kilo\meter} is large enough that upper-boundary effects do not affect the solution.
The model is integrated forward in time using an implicit timestepping scheme with a timestep of one day.

The 1D BL model yields an excellent approximation of the full 1D PG solution (Fig.~\ref{fig:slope_profiles}).
The interior dynamics match the interior of the full solution, and although the BL model only explicitly computes the interior evolution, the BL correction computed offline from~\eqref{eq:1d-bl-sol} is very accurate.
The match is trivial when~$\mu = 1$ and~$\varrho = \num{e-3}$, because the shallow slope leads to a relatively weak BL transport, and thus the advective modification to the buoyancy flux in~\eqref{eq:1d-evolution-interior} and~\eqref{eq:1d-interior-bc} is negligible.
The interior system is then nearly identical to the full one, with diffusion dominating the dynamics.
The case where~$\varrho = 0.5$, in contrast, is a more trying test of the 1D BL theory.
The BL transport in this case is an order of magnitude larger than before, leading to enhanced stratification in the BL.
This is properly captured in the BL model, with the interior stratification reaching about~\SI{0.4e-6}{\per\second\squared} at the bottom and the BL correction bringing it smoothly to zero.

The assumption of 1D dynamics breaks down as soon as lateral variations in the slope are allowed, but we can anticipate the upcoming 2D results using intuition derived from the above 1D theory.
Equation~\eqref{eq:1d-bl-transport} gives an explicit expression for the BL transport in 1D depending on the local slope angle~$\theta$ and buoyancy gradient across the slope~$N^2 \tan \theta$.
In 2D, these inputs are spatially dependent, with horizontal buoyancy gradients also varying in time as part of the interior dynamics.
Local 1D theory would thus predict convergences and divergences in BL transport, generating BL--interior mass exchange (Fig.~\ref{fig:BL_sketch}b).
This leads to a more complex picture in 2D, with interior dynamics feeding back onto the BL, as we will see in the following section.

\section{Two-dimensional boundary layer theory}\label{s:2d}

In this section, we extend the 1D BL theory to the 2D PG equations in terrain-following coordinates.
We first derive the 2D BL equations and then apply them to idealized numerical simulations.

\subsection{Boundary layer theory}

\begin{figure}
    \centering
    \includegraphics[width=19pc]{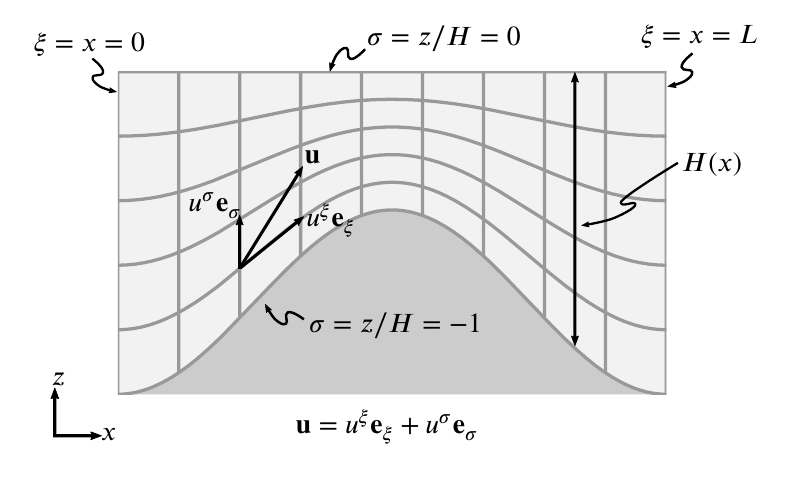}
    \caption{Sketch of terrain-following coordinates used in 2D BL theory.
    The covariant basis vector of coordinate~$j$ is denoted by~$\vec{e}_j$, and the corresponding contravariant component of the velocity vector is denoted by~$u^j$, such that $\vec{u} = u^j \vec{e}_j$ (summation implied).}
    \label{fig:TF_coords}
\end{figure}

In 2D, the interaction between the BL and interior is more interesting because, in addition to the BL advection imposing a buoyancy flux on the interior, variations in the BL transport produce mass exchange with the interior \citep[e.g.,][]{phillips_experiment_1986,mcdougall_dianeutral_1989,kunze_turbulent_2012,dell_diffusive_2015,ledwell_comment_2018,holmes_ridges_2018}. 
The BL theory generalizes from 1D to 2D and brings these physics into clearer focus.

Applying the BL theory to the 2D PG equations is most easily done in terrain-following coordinates:
\begin{equation}
    \xi = x, \quad \eta = y, \quad \sigma = \frac{z}{H},
\end{equation} 
where~$H(x)$ is the fluid depth (Fig.~\ref{fig:TF_coords}).
Under this transformation, derivatives in~$(x, z)$ space become 
\begin{equation}
    \pder{}{x} = \pder{}{\xi} - \frac{\sigma \partial_x H}{H}\pder{}{\sigma} \quad \text{and} \quad \pder{}{z} = \frac{1}{H}\pder{}{\sigma},
\end{equation}
and the contravariant velocity components are
\begin{equation}
    u^\xi = u^x, \quad u^\eta = u^y, \quad \text{and} \quad u^\sigma = \frac{1}{H}\left(u^z - \sigma\pder{H}{x}u^x\right),
\end{equation}
assuming no variations in~$\eta$ (see appendix~B of \citet{callies_dynamics_2018} for more details).
The 2D PG equations in terrain-following coordinates are then
\begin{align}
    -fu^\eta &= -\pder{p}{\xi} + \sigma \pder{H}{x} b + \frac{1}{H^2}\pder{}{\sigma}\left(\nu \pder{u^\xi}{\sigma}\right),\label{eq:2d-xi}\\
    fu^\xi &= \frac{1}{H^2}\pder{}{\sigma}\left(\nu \pder{u^\eta}{\sigma}\right),\label{eq:2d-eta}\\
    \frac{1}{H}\pder{p}{\sigma} &= b,\label{eq:2d-sigma}\\
    \pder{}{\xi}\left(H u^\xi\right) + \pder{}{\sigma}\Big(H u^\sigma\Big) &= 0,\label{eq:2d-continuity}\\
    \pder{b}{t} + u^\xi \pder{b}{\xi} + u^\sigma \pder{b}{\sigma} &= \frac{1}{H^2} \pder{}{\sigma} \left( \kappa \pder{b}{\sigma} \right),\label{eq:2d-buoyancy}
\end{align}
where~$p$ is the pressure divided by a reference density.
The boundary conditions are again an insulating and no-slip bottom, $\partial_\sigma b = 0$ and~$u^\xi = u^\eta = 0$ at~$\sigma = -1$; a constant-flux and free-slip top~$H^{-1} \partial_\sigma b = N^2$ and~$\partial_\sigma u^\xi = \partial_\sigma u^\eta = 0$ at~$\sigma = 0$; and no normal flow across both boundaries, $u^\sigma = 0$ at~$\sigma = -1$ and~$\sigma = 0$.
We neglect horizontal turbulent fluxes, consistent with the assumption of a small aspect ratio if the turbulence is close to isotropic.
This is in contrast with some other PG models, which employed horizontal diffusion terms to satisfy the no-normal-flow condition at vertical side-walls \citep[e.g.,][]{verdieere_mean_1986,samelson_simple_1997}.

As before, we express the momentum equations~\eqref{eq:2d-xi} to~\eqref{eq:2d-continuity} as one streamfunction inversion.
We define~$\chi(\xi, \sigma)$ such that the continuity equation~\eqref{eq:2d-continuity} is automatically satisfied:
\begin{equation}
    H u^\xi = \pder{\chi}{\sigma} \quad \text{and} \quad H u^\sigma = -\pder{\chi}{\xi}.
\end{equation}
Integrating~\eqref{eq:2d-eta} from some level to~$\sigma = 0$, we obtain
\begin{equation}\label{eq:2d-ueta-from-chi}
    \frac{1}{H}\pder{u^\eta}{\sigma} = \frac{f}{\nu}(\chi - U^\xi),
\end{equation}
as in equation~\eqref{eq:1d-ueta-from-chi}.
Here, $U^\xi = \int_{-1}^0 H u^\xi \; d\sigma$ is the vertically integrated transport, a constant in~$\xi$ by continuity.
Combining~$H^{-1} \partial_\sigma$ of~\eqref{eq:2d-xi} and~$\partial_\xi$ of~\eqref{eq:2d-sigma} and substituting~$H^{-1} \partial_\sigma u^\eta$ from~\eqref{eq:2d-ueta-from-chi} yields the streamfunction inversion equation similar to 1D:
\begin{equation}\label{eq:2d-inversion}
    \frac{1}{H^4}\pder{^2}{\sigma^2}\left(\nu \pder{^2\chi}{\sigma^2}\right) + \frac{f^2}{\nu}(\chi - U^\xi) = \pder{b}{\xi} - \frac{\sigma}{H} \pder{H}{x} \pder{b}{\sigma}.
\end{equation}
The boundary conditions are similar to the 1D case but for a finite domain: $\chi = 0$ and~$\partial_\sigma \chi = 0$ at~$\sigma = -1$ and~$\chi = U^\xi$ and~$\partial^2_\sigma \chi = 0$ at~$\sigma = 0$.

Splitting $b$ and~$\chi$ into BL and interior contributions and neglecting the fourth-order term in \eqref{eq:2d-inversion} in the interior as before, the interior inversion reads
\begin{equation}\label{eq:2d-inversion-interior}
    \frac{f^2}{\nu}\chi\I = \pder{b\I}{\xi} - \frac{\sigma}{H}\pder{H}{x}\pder{b\I}{\sigma} = \pder{b\I}{x},
\end{equation}
setting~$U^\xi = 0$ as implied by a configuration that is symmetric in~$x$ (see appendix~A for the $U^\xi \neq 0$ case).
The circulation in the~$x$--$z$ plane is simply proportional to the buoyancy gradient in~$x$.
The interior buoyancy evolution is given by
\begin{equation}\label{eq:2d-evolution-interior}
    \pder{b\I}{t} + \frac{1}{H} \left( \pder{\chi\I}{\sigma} \pder{b\I}{\xi} - \pder{\chi\I}{\xi} \pder{b\I}{\sigma} \right) = \frac{1}{H^2} \pder{}{\sigma} \left( \kappa \pder{b\I}{\sigma} \right).
\end{equation}
The BL physics appear in the boundary condition on the interior buoyancy field.
The BL buoyancy budget, assuming a quasi-steady state and a slowly varying interior buoyancy field, is
\begin{equation}
    \frac{1}{H} \pder{\chi\B}{\sigma} \pder{b\I}{\xi} = \frac{1}{H^2} \pder{}{\sigma} \left( \kappa \pder{b\B}{\sigma} \right),
    \label{eq:2d-bl-buoyancy}
\end{equation}
with $\partial_\xi b\I$ evaluated at $\sigma = -1$.
The neglected advection terms are smaller by a factor $(qH)^{-1} \ll 1$ than the terms retained in~\eqref{eq:2d-bl-buoyancy}.
This is because the boundary conditions enforce that $\chi\B \sim \chi\I$ and $\partial_\sigma b\B \sim \partial_\sigma b\I$, such that $\partial_\sigma \chi\B \sim (qH) \partial_\sigma \chi\I$ and $b\B \sim (qH)^{-1} b\I$ (see section~\ref{s:asymptotics} for more detail).
Vertically integrating \eqref{eq:2d-bl-buoyancy} across the BL and applying the boundary conditions $\chi\I + \chi\B = 0$ and $\partial_\sigma b\I + \partial_\sigma b\B = 0$ at $\sigma = -1$, as well as decay conditions for $\chi\B$ and $\partial_\sigma b\B$, yields
\begin{equation}\label{eq:2d-interior-bc}
    \chi\I \pder{b\I}{\xi} = \frac{\kappa}{H}\pder{b\I}{\sigma} \quad \text{at} \quad \sigma = -1.
\end{equation}
Substituting this bottom boundary condition for the interior into the interior inversion~\eqref{eq:2d-inversion-interior}, we again arrive at an explicit formula for this BL transport:
\begin{equation}\label{eq:2d-bl-transport}
    \chi\I = \frac{\kappa}{\pder{H}{x}} \frac{ \frac{\mu}{f^2} \pder{H}{x} \pder{b\I}{\xi}}{1 - \frac{\mu}{f^2} \pder{H}{x} \pder{b\I}{\xi}} = \frac{\frac{\nu}{f^2} \pder{b\I}{\xi}}{1 - \frac{\mu}{f^2} \pder{H}{x} \pder{b\I}{\xi}} \quad \text{at} \quad \sigma = -1.
\end{equation}
This is the generalization of the 1D result~\eqref{eq:1d-bl-transport}: $-\partial_x H$ is analogous to the local slope~$\tan \theta$ and~$\partial_\xi b\I$ now takes the place of the previously constant cross-slope buoyancy gradient~$N^2 \tan\theta$.
Note that this expression is again well-behaved in the limit of small slopes ($\partial_x H \to 0$) and thus gives a globally valid expression for the BL transport and of the mass exchange $H u^\sigma\I = -\partial_\xi \chi\I$ at $\sigma = -1$ between the BL and the interior.

As in 1D, we can now explicitly describe contributions to the interior along-slope flow from thermal wind in the interior and a contribution from shear in the BL.
Combining~\eqref{eq:2d-ueta-from-chi} and~\eqref{eq:2d-inversion-interior} yields the thermal-wind balance
\begin{equation}
    \frac{1}{H} \pder{u^\eta\I}{\sigma} = \frac{1}{f} \pder{b\I}{x},
\end{equation}
which, upon integration in the vertical, becomes
\begin{equation}\label{eq:2d-ueta-interior}
    u^\eta\I(\sigma) = -\frac{f \chi\I(-1)}{\nu(-1) q} + \frac{H}{f} \int_{-1}^\sigma \pder{b\I}{x}(\tilde \sigma) \; d\tilde\sigma.
\end{equation}
The first term again represents the BL contribution~$u^\eta\I = -u^\eta\B$ at~$\sigma = -1$, which may be computed directly from the BL solution
\begin{equation}\label{eq:2d-bl-correction}
    \chi\B = -\chi\I e^{-q H (\sigma + 1)} [ \cos q H (\sigma + 1) + \sin q H (\sigma + 1) ],
\end{equation}
similar to~\eqref{eq:1d-bl-sol}.
Here~$q$ can still be written in the same form as in~\eqref{eq:q} but with a generalized slope Burger number~$\varrho = -\partial_x H \partial_\xi b\I(-1) / f^2$, which varies in the horizontal.
Equation~\eqref{eq:2d-ueta-interior} has the same form as~\eqref{eq:1d-ueta-interior}, except that cross-slope buoyancy gradients can now contribute to the thermal-wind term.

In 2D, we again find that the interior solution experiences a buoyancy flux due to the cross-slope advection by the BL transport. 
In contrast to the 1D case, however, both the BL transport given by~\eqref{eq:2d-bl-transport} and the cross-slope buoyancy gradient~$\partial_\xi b\I$ may vary in time and space (Fig.~\ref{fig:BL_sketch}b).
Convergence in the BL transport then drives mass injection into the interior, further altering~$\partial_\xi b\I$ and continuing the feedback process.

It is worth noting that BL theory can also be applied to a passive tracer, not just buoyancy.
The interior tracer concentration would have a similar effective boundary condition capturing transport by BL flow.
The interior tracer equation should also include a representation of along-isopycnal stirring \citep{redi_oceanic_1982}.

\subsection{Examples}

We now illustrate these theoretical results using numerical simulations over idealized topographies.
We solve the full 2D PG system~\eqref{eq:2d-buoyancy} and~\eqref{eq:2d-inversion} and the 2D BL PG system~\eqref{eq:2d-inversion-interior} and~\eqref{eq:2d-evolution-interior} using numerical methods and model parameters similar to the 1D case described above.
The mixing profile is now written as
\begin{equation}
    \kappa = \kappa_0 + \kappa_1 e^{-(z + H)/h},
\end{equation}
following the bottom topography.
First, we study spin up over an idealized azimuthally symmetric seamount with constant initial stratification.
We then analyze spin up over an idealized mid-Atlantic ridge with both constant and exponentially varying initial stratification.
As in the 1D spin up experiments, the simulations all start with flat isopycnals and no flow.
The circulation that emerges is powered by the potential-energy source $\kappa \partial_z b$ integrated over the domain.

\subsubsection{Idealized seamount}

\begin{figure}
    \centering
    \includegraphics[width=19pc]{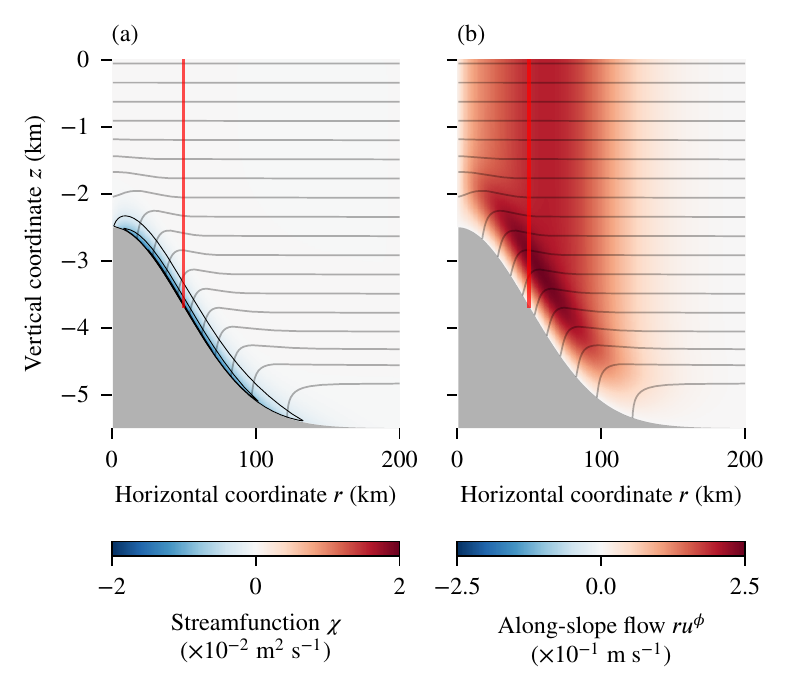}
    \caption{Flow fields in a simulation of mixing-generated PG spin up over an idealized 2D seamount.
    Shown are (a)~the streamfunction~$\chi$ (shading and black contours) with positive values indicating counter-clockwise and negative values indicating clockwise flow and (b)~the along-slope flow~$u^y = u^\eta$.
    The solution is shown after 20~years of spin up.
    The gray curves show isopycnals, and the red vertical lines show where 1D profiles are examined in Fig.~\ref{fig:seamount_profiles}.
    }
    \label{fig:seamount}
\end{figure}

\begin{figure*}
    \centering
    \includegraphics[width=27pc]{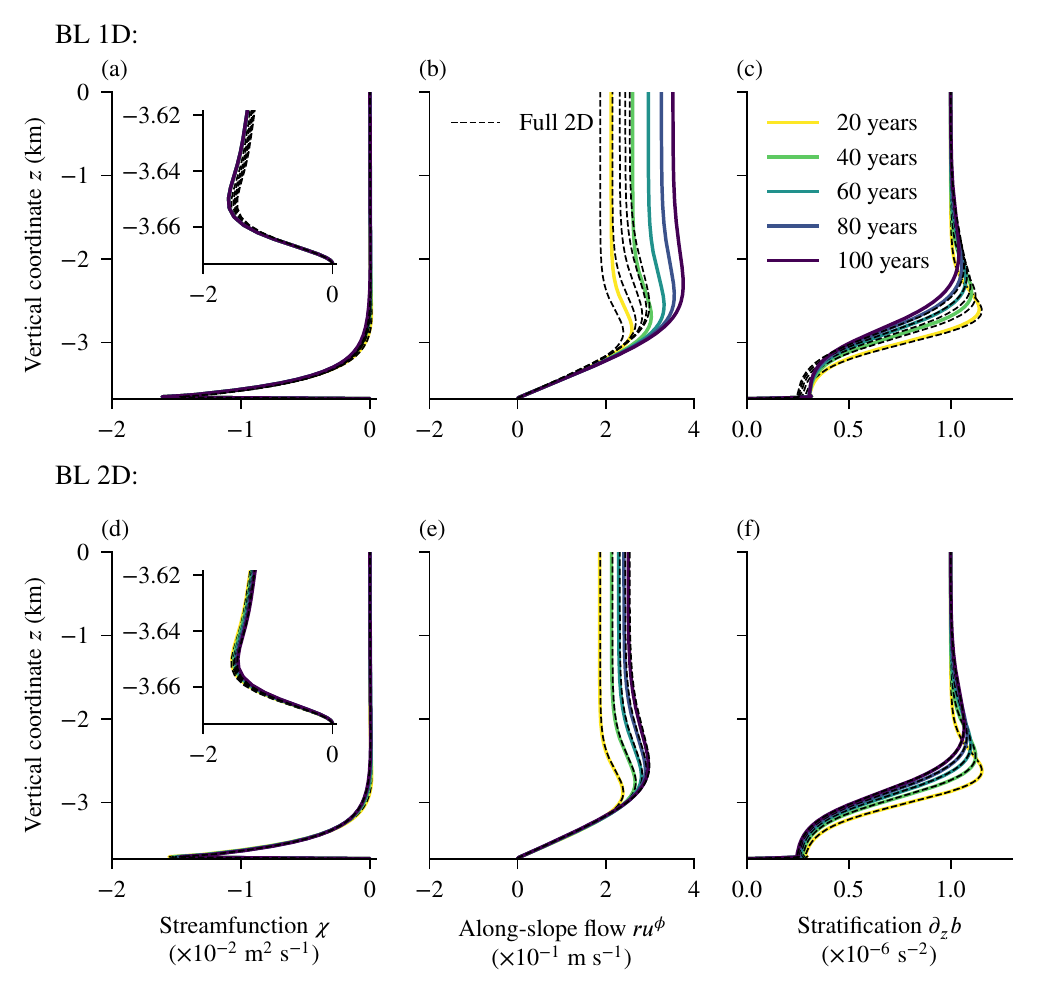}
    \caption{Comparison of the 1D and 2D BL solutions with full 2D PG mixing-generated spin up over a seamount.
    Profiles are taken at the steepest slope on the seamount (red lines in Fig.~\ref{fig:seamount}).
    Shown are the (a),~(d)~streamfunction~$\chi$, (b),~(e)~along-slope flow~$u^y = u^\eta$, and (c),~(f)~stratification~$\partial_z b$.
    The insets of (a) and (d) show the streamfunction~$\chi$ in the bottom~\SI{50}{\meter}, showcasing the BL correction.
    The 1D BL solution is a decent approximation to the flow, but the cross-slope variations considered in the 2D BL theory allow it to better match the full 2D solution in this high slope Burger number regime.
    }    
    \label{fig:seamount_profiles}
\end{figure*}

The topography of the abyssal ocean has a range of slopes.
Seamounts, for instance, can reach slope Burger numbers of order 10 or more and have received some attention regarding their role in the abyssal overturning circulation \citep[e.g.,][]{mcdougall_dianeutral_1989,mcdougall_abyssal_2017,ledwell_comment_2018,holmes_ridges_2018}.
The 1D BL theory [equation~\eqref{eq:1d-evolution-interior}] is sensitive to the slope Burger number, with a steeper slope leading to a larger modification of the diffusive buoyancy flux by advection.
At the same time, the 2D BL theory shows that horizontal variations in this slope lead to gradients in BL transport that are not taken into account by the 1D theory.
In this section, we therefore compare both 1D and 2D BL solutions to the full 2D PG flow over a seamount.

Similar to the analysis in \citet{ledwell_comment_2018}, we consider an azimuthally symmetric Gaussian seamount in axisymmetric coordinates (Fig.~\ref{fig:seamount}).
On an~$f$-plane, the flow is invariant under rotation about the center of the seamount, allowing us to fully describe the flow using 2D theory (see appendix~B).
The depth of the seafloor as a function of distance~$r$ from the symmetry axis is given by
\begin{equation}
    H(r) = H_0 - A \exp \left( -\frac{r^2}{2\ell^2} \right),
\end{equation}
where the maximum depth is~$H_0 = \SI{5.5}{\kilo\meter}$, the height of the seamount is~$A = \SI{3}{\kilo\meter}$, the width of the seamount is~$\ell = \SI{50}{\kilo\meter}$, and the width of the domain is $L = \SI{200}{\kilo\meter}$.
We assume no flow at~$r = 0$ and allow the flow to evolve freely at~$r=L$, consistent with our assumption that horizontal diffusion may be neglected.
In the horizontal, the grid has an even spacing of about~\SI{0.8}{\kilo\meter}.
As in the 1D models, we use Chebyshev nodes in the vertical when solving the full 2D PG equations (with a near-bottom resolution of about~\num{e-5} in~$\sigma$-space) and uniform grid spacing for the 2D BL equations (with a resolution of about~\num{e-3} in~$\sigma$-space).
We initialize the model at rest with a constant stratification~$b = N^2 z$ and use a mixed implicit--explicit time integration scheme with a timestep of one day.

At the steepest point on the seamount ($r = \SI{50}{\kilo\meter}$, red lines in Fig.~\ref{fig:seamount}), the slope Burger number~$\varrho$ is order unity.
The 1D BL solution applied at this position over-predicts the stratification in the bottom~\SI{500}{\meter} and under-predicts it above (Fig.~\ref{fig:seamount_profiles}).
This leads to errors in the predicted interior along-slope flow, which can be understood from~\eqref{eq:1d-ueta-interior} and~\eqref{eq:2d-ueta-interior}: even subtle changes in the buoyancy field can lead to substantial impacts on~$u^\eta\I$ after being integrated throughout the column.
The 1D BL solution's buoyancy field differs from that of the 2D solution because its secondary circulation, enforced simply by a transport constraint, is stronger.
This is due to the lack of a two-way feedback in 1D; the BL cannot exchange mass with the interior and the induced changes in the interior do not reduce the BL transport. 
The 2D BL theory, in contrast, captures these physics and agrees well with the full 2D model.
This confirms that the 2D BL equations are capable of fully capturing 2D PG spin up, even in regimes with relatively large variations in local slope.

\subsubsection{Exponential background stratification}\label{sss:N2exp}

\begin{figure*}
    \centering
    \includegraphics[width=33pc]{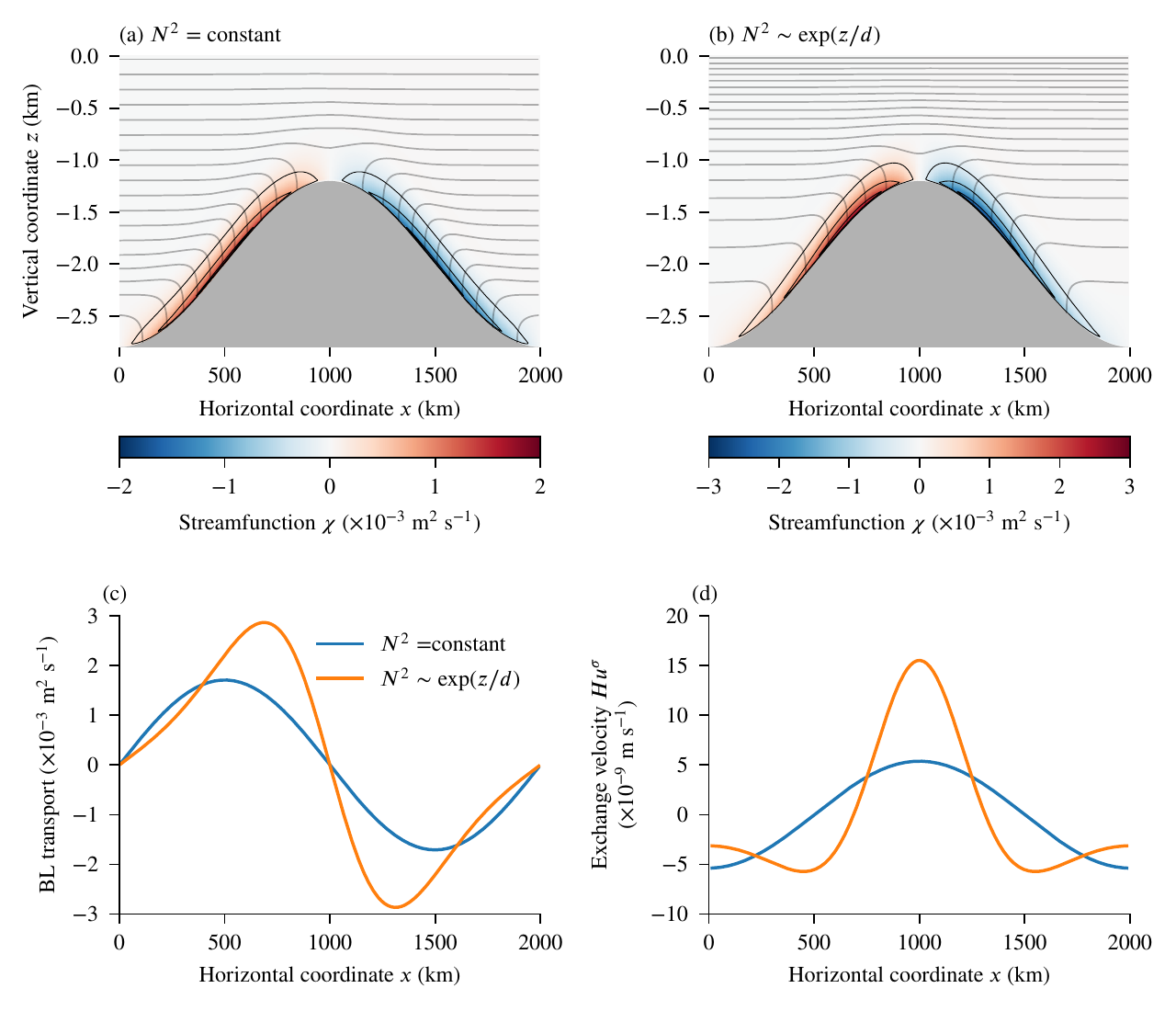}
    \caption{Simulations of mixing-generated PG spin up over an idealized 2D mid-ocean ridge with varying initial stratifications.
    Shown are the streamfunctions~$\chi$ (shading and black contours) with positive values indicating counter-clockwise and negative values indicating clockwise flow for simulations with (a)~constant initial stratification and (b)~exponential initial stratification (isopycnals in gray).
    For each simulation, we show (c)~the BL transport~$U^\xi\B$ computed from equation~\eqref{eq:2d-bl-transport} and (d)~the resulting exchange velocity~$Hu^\sigma = -\partial_\xi U^\xi\B$.
    The solutions are shown after three years of spin up.
    The gradient in stratification across the ridge facilitates larger exchange velocities at the peak and flanks.}
    \label{fig:ridge_exp_strat}
\end{figure*}

The simulations presented so far were initialized with a constant background stratification.
In the real ocean, the stratification varies significantly in the vertical, often decreasing close to exponentially with depth \citep[e.g.,][]{munk_abyssal_1966}.
A number of studies have attempted to discern how this may shape the abyssal circulation, often qualitatively arguing that variations in stratification across slopes must lead to gradients in BL transports, inducing BL--interior exchange \citep[e.g.,][]{phillips_experiment_1986,salmun_two-dimensional_1991}.
Quantitative explanations of this process, however, have remained complicated and opaque at best.
A major benefit of the BL theory framework built up here is that it provides concise expressions for the BL transport in terms of interior variables, allowing us to reason about how varying background stratification might impact the abyss with minimal mathematical gymnastics.

Let us consider an idealized mid-Atlantic ridge, following previous studies of mixing-generated spin up in the abyss \citep[e.g.,][\citetalias{peterson_rapid_2022}]{ruan_mixing-driven_2020,drake_abyssal_2020}.
The depth of the 2D ridge is given by
\begin{equation}
    H(x) = H_0 + A \cos \left( \frac{2\pi x}{L} \right),
\end{equation}
where the mean depth is~$H_0 = \SI{2}{\kilo\meter}$, the amplitude is~$A = \SI{800}{\meter}$, and the width is~$L = \SI{2000}{\kilo\meter}$ (Fig.~\ref{fig:ridge_exp_strat}).
At the steepest point on the ridge, the slope Burger number~$\varrho$ is approximately~\SI{2e-3}, typical of the mid-Atlantic ridge.
We apply periodic boundary conditions at~$x = 0$ and~$x = L$ and use a constant horizontal grid spacing of about~\SI{8}{\kilo\meter}.
The vertical grid spacing is as before.
We run one simulation with constant initial stratification as before and one initialized with an exponential stratification: $\partial_z b \propto e^{z/d}$.
We set the decay scale to~$d = \SI{1000}{\meter}$ and choose the proportionality constant such that the bottom stratification at the center of the ridge flank matches that of the simulation with constant~$N^2 = \SI{e-6}{\per\second\squared}$.
We again use a mixed implicit--explicit timestepping scheme, this time with a timestep of 10~days, enabled by the much weaker advective terms.

The circulation in the case with exponential initial stratification is stronger and more confined to the peak of the ridge compared to the case with constant initial stratification (Fig.~\ref{fig:ridge_exp_strat}a,b).
This is better understood by the explicit formula for 2D BL transport derived in the previous subsection.
Evaluating equation~\eqref{eq:2d-bl-transport} for these simulations, we see that the BL transport is enhanced at the peak of the ridge with exponential background stratification (Fig.~\ref{fig:ridge_exp_strat}c).
For the small slopes in this simulation, equation~\eqref{eq:2d-bl-transport} reduces to 
\begin{equation}
    \chi\I \approx \frac{\nu}{f^2} \pder{b\I}{\xi} \quad \mathrm{at} \quad \sigma = -1.
\end{equation}
In the case with constant stratification, the initial cross-slope buoyancy gradient is proportional to $-\partial_x H$ and does not change appreciably with time, explaining the sinusoidal BL transport.
For exponential stratification, in contrast, we have~$\partial_\xi b\I \propto -e^{-H/d} \partial_x H$, which is enhanced at shallower depths.
As a result, the exchange velocity 
\begin{equation}
    Hu^\sigma = -\partial_\xi \chi\I \approx - \frac{\nu}{f^2} \pder{^2 b\I}{\xi^2} \quad \mathrm{at} \quad \sigma = -1
\end{equation}
is also enhanced for the case with exponential stratification (Fig.~\ref{fig:ridge_exp_strat}d).
In both cases, $\partial_\xi b\I$ does not evolve much in the first three years, so the exchange does not either.
The BL theory enables us to easily and quantitatively understand this behavior.

\section{Asymptotic theory}\label{s:asymptotics}

In the previous sections, we derived the BL equations somewhat heuristically, glossing over some detail of the underlying asymptotics.
In this section, we present a more rigorous derivation of the BL theory that justifies the claims in the previous sections and sheds light on the asymptotic orders of the various components of the flow.
The casual reader should note that the contents of this section are not required to understand the main results of the paper.

We show below that, in both 1D and 2D, the cross-slope flow is of lower order than the along-slope flow in the interior, aligning with our intuition from the examples above.
The interior flow evolves on a slow timescale driven by diffusion and second-order advection of the leading-order buoyancy in the interior.
The BL flow is of first order, in between the orders of the interior along- and cross-slope flows.
If the transport is constrained to zero, this implies that the leading-order interior flow vanishes at the bottom.
These results do not generally hold in 3D, but we leave this generalization to future work.

\subsection{One-dimensional asymptotics}

To begin the formal derivation of the 1D BL equations, we first nondimensionalize the 1D equations~\eqref{eq:1d-xi}--\eqref{eq:1d-U} in order to isolate the key parameters in the problem.
We define characteristic scales for the vertical coordinate, velocities, and mixing coefficients such that
\begin{equation}\label{eq:1d-scales}
    \zeta \sim H_0, \quad u^\xi, u^\eta \sim U, \quad \nu \sim \nu_0, \quad \mathrm{and} \quad \kappa \sim \kappa_0,
\end{equation}
where $\nu_0$ and $\kappa_0$ are characteristic values of $\nu$ and~$\kappa$.
We assume that the pressure and buoyancy terms in~\eqref{eq:1d-xi} scale with the Coriolis term and that the buoyancy perturbation scales with the background buoyancy scale:
\begin{equation}\label{eq:1d-scales-P-b}
    \pder{P}{x} \sim fU \quad \mathrm{and} \quad b' \sim \frac{f U}{\tan\theta} = N^2 H_0.
\end{equation}
Assuming an advective timescale, so that
\begin{equation}\label{eq:1d-scales-time}
    t \sim \frac{H_0}{U \tan\theta} = \frac{f}{N^2 \tan^2\theta},
\end{equation}
then yields the nondimensional 1D equations
\begin{align}
    -u^\eta &= -\pder{P}{x} + b' + \varepsilon^2 \pder{}{\zeta}\left( \nu \pder{u^\xi}{\zeta} \right),\label{eq:1d-xi-nd}\\
    u^\xi &= \varepsilon^2 \pder{}{\zeta}\left( \nu \pder{u^\eta}{\zeta} \right),\label{eq:1d-eta-nd}\\
    \mu \varrho \left( \pder{b'}{t} + u^\xi \right) &= \varepsilon^2 \pder{}{\zeta}\left[ \kappa \left( 1 + \pder{b'}{\zeta} \right) \right],\label{eq:1d-b-nd}\\
    \int_0^\infty u^\xi \; d\zeta &= U^\xi,\label{eq:1d-U-nd}
\end{align}
where all variables are redefined to their scaled versions.
The nondimensional parameters for the 1D problem are thus the Ekman number~$\varepsilon^2 = \nu_0 / f H_0^2$, the Prandtl number~$\mu = \nu_0 / \kappa_0$, and the slope Burger number~$\varrho = N^2 \tan^2\theta / f^2$, although $\mu$ and $\varrho$ only appear as a product, so $\mu \varrho$ can be considered a single parameter.
The reason for defining the Ekman number as~$\varepsilon^2$ will become clear in the BL analysis below.

To develop the asymptotic theory, we assume the scaling $\varepsilon \ll 1$ and~$\mu \varrho \sim 1$.
While the Burger number is typically small in the abyss, the turbulent Prandtl number may be large if momentum fluxes by baroclinic eddies are taken into account.
If instead $\mu \varrho \ll 1$, buoyancy advection is negligible in the BL, and the theory developed with $\mu \varrho \sim 1$ remains accurate (Fig.~\ref{fig:slope_profiles}a).

We begin with the interior and expand all variables in $\varepsilon^2$: $u^\xi\I = u^\xi\Izero + \varepsilon^2 u^\xi\Itwo + \dots$, etc.
This expansion into even powers of $\varepsilon$ is sufficient because $\varepsilon$ only appears as $\varepsilon^2$ in the interior equations.
The $O(1)$ interior flow then satisfies
\begin{align}
    -u^\eta\Izero &= -\pder{P_0}{x} + b'\Izero,\label{eq:1d-xi-I0}\\
    u^\xi\Izero &= 0,\\
    \pder{b'\Izero}{t} &= 0.
\end{align}
At this order, the interior along-slope flow is in balance with the barotropic pressure gradient and the projection of the buoyancy perturbation, and the interior cross-slope flow is zero.
The $O(1)$ buoyancy equation is trivial, implying that the interior buoyancy evolution is slow compared to the advective timescale assumed in the scaling.

To obtain the evolution of the~$O(1)$ interior buoyancy, we need to go to $O(\varepsilon^2)$ and also expand the time coordinate, $\partial_t = \partial_{t_0} + \varepsilon^2 \partial_{t_2} + \dots$
Higher-order buoyancy terms inherit the slow evolution from the low orders, so $\partial_{t_0} b'\Itwo = 0$.
The buoyancy equation~\eqref{eq:1d-b-nd} at~$O(\varepsilon^2)$ is then
\begin{equation}\label{eq:1d-b-I0}
    \mu \varrho \left( \pder{b'\Izero}{t_2} + u^\xi\Itwo \right) = \pder{}{\zeta} \left[ \kappa \left( 1 + \pder{b'\Izero}{\zeta} \right) \right].
\end{equation}
This implies that advection and turbulent diffusion operate on a slow time~$t_2$.
Since the~$O(1)$ and~$O(\varepsilon)$ interior cross-slope flows are zero, the dominant buoyancy advection is by the second-order flow in the interior, given by~\eqref{eq:1d-eta-nd} at~$O(\varepsilon^2)$:
\begin{equation}\label{eq:1d-eta-I0}
    u^\xi\Itwo = \pder{}{\zeta}\left( \nu \pder{u^\eta\Izero}{\zeta} \right).
\end{equation}
Equations~\eqref{eq:1d-xi-I0}, \eqref{eq:1d-b-I0}, and~\eqref{eq:1d-eta-I0} comprise the leading-order interior dynamics.
They can be expressed in terms of the streamfunction~$\chi\I$, whose leading non-zero component is~$\chi\Itwo$, recovering~\eqref{eq:1d-inversion-interior} and~\eqref{eq:1d-evolution-interior} above (assuming $U^\xi = 0$).
The interior along-slope flow can be obtained by integrating the thermal-wind balance $\partial_\zeta u^\eta\Izero = -\partial_\zeta b'\Izero$, which follows from a $\zeta$-derivative of~\eqref{eq:1d-xi-I0}:
\begin{equation}
    u^\eta\Izero = u^\eta\Izero(0) - \left[ b'\Izero - b'\Izero(0) \right].
\end{equation}
The integration constant $u^\eta\Izero(0)$ must be determined from the BL correction.
If the transport constraint is $U^\xi = 0$, one finds that $u^\eta\Izero(0) = 0$.

In the thin bottom BL, $\zeta$-derivatives are enhanced, elevating the diffusion terms in~\eqref{eq:1d-xi-nd}--\eqref{eq:1d-b-nd} to $O(1)$.
Given that the BL thickness scales with $\varepsilon$, we assume the BL variables to depend on the re-scaled vertical coordinate $\bar\zeta = \zeta/\varepsilon$, with which~$\partial_{\zeta} = \varepsilon^{-1} \partial_{\bar\zeta}$.
The nondimensional BL equations are then
\begin{align}
    -u^\eta\B &= b'\B + \pder{}{\bar\zeta}\left( \nu \pder{u^\xi\B}{\bar\zeta} \right),\\
    u^\xi\B &= \pder{}{\bar\zeta}\left( \nu \pder{u^\eta\B}{\bar\zeta} \right),\\
    \mu \varrho \left( \pder{b'\B}{t} + u^\xi\B \right) &= \pder{}{\bar\zeta}\left( \kappa \pder{b'\B}{\bar\zeta} \right).
\end{align}
Crucially, the insulating bottom boundary condition picks up a factor of~$\varepsilon^{-1}$ after this re-scaling:
\begin{equation}
    1 + \pder{b'\I}{\zeta} = -\frac{1}{\varepsilon} \pder{b'\B}{\bar\zeta} \quad \mathrm{at} \quad \zeta = 0.
\end{equation}
This factor of $\varepsilon^{-1}$ means that we need an $O(\varepsilon)$ BL buoyancy to absorb the $O(1)$ interior buoyancy flux into the BL.
We thus expand the BL variables in terms of~$\varepsilon$ rather than $\varepsilon^2$.
We immediately find that the $O(1)$ BL buoyancy flux must vanish at the bottom: $\partial_{\bar\zeta} b'\Bzero = 0$.
In the case with zero net transport ($U^\xi = 0$), this condition, along with the boundary conditions on the flow~$u^\xi\Bzero = 0$ and $u^\eta\Bzero = -u^\eta\Izero$ at $\bar\zeta = 0$, forces the $O(1)$ BL flow to vanish and the $O(1)$ interior along-slope flow to go to zero at the bottom, consistent with the examples shown in Fig.~\ref{fig:slope_profiles} (see appendix~A for the $U^\xi \neq 0$ case).
The BL flow instead comes in at~$O(\varepsilon)$, in between the orders of the interior along- and cross-slope flows.
This $O(\varepsilon)$ BL flow satisfies
\begin{align}
    -u^\eta\Bone &= b'\Bone + \pder{}{\bar\zeta}\left( \nu \pder{u^\xi\Bone}{\bar\zeta} \right),\\
    u^\xi\Bone &= \pder{}{\bar\zeta}\left( \nu \pder{u^\eta\Bone}{\bar\zeta} \right),\\
    \mu \varrho u^\xi\Bone &= \pder{}{\bar\zeta}\left( \kappa \pder{b'\Bone}{\bar\zeta} \right),\label{eq:1d-b-B1}
\end{align}
with the bottom boundary conditions $\partial_{\bar\zeta} b'\Bone = -(1 + \partial_\zeta b'\Izero)$, $u^\xi\Bone = 0$, and $u^\eta\Bone = 0$.
The tendency term~$\partial_{t_0} b'\Bone$ is dropped because the interior does not evolve on this timescale, so the BL will not either.
These BL equations are equivalent to~\eqref{eq:1d-bl-buoyancy} and~\eqref{eq:1d-bl-equation}.

This more rigorous derivation of the 1D BL equations clarifies the asymptotic orders of the various components of the flow.
The leading-order contributions are $O(\varepsilon^2)$ for the interior cross-slope flow, $O(1)$ for the interior along-slope flow, and $O(\varepsilon)$ for both components of the BL flow.
Buoyancy does not have an $O(1)$ BL correction---only its derivative does.

\subsection{Two-dimensional asymptotics}

The 2D asymptotics follow in much the same way as in 1D.
We again nondimensionalize the equations of motion~\eqref{eq:2d-xi}--\eqref{eq:2d-buoyancy}, setting characteristic scales equivalent to~\eqref{eq:1d-scales}--\eqref{eq:1d-scales-time}:
\begin{multline}
    \xi \sim L, \quad u^\xi, u^\eta \sim U, \quad u^\sigma \sim \frac{U}{L}, \quad H \sim H_0, \quad \nu \sim \nu_0, \quad \kappa \sim \kappa_0,\\
    \quad p \sim U f L, \quad b \sim \frac{f U L}{H_0} = N^2 H_0, \quad t \sim \frac{L}{U}.
\end{multline}
We then arrive at the nondimensional 2D PG equations
\begin{align}
    -u^\eta &= -\pder{p}{\xi} + \sigma \pder{H}{x} b + \frac{\varepsilon^2}{H^2} \pder{}{\sigma}\left(\nu \pder{u^\xi}{\sigma}\right),\\
    u^\xi &= \frac{\varepsilon^2}{H^2} \pder{}{\sigma}\left(\nu \pder{u^\eta}{\sigma}\right),\\
    \frac{1}{H} \pder{p}{\sigma} &= b,\\
    \pder{}{\xi}\left(H u^\xi\right) + \pder{}{\sigma}\Big(H u^\sigma\Big) &= 0,\\
    \mu \varrho \left( \pder{b}{t} + u^\xi \pder{b}{\xi} + u^\sigma \pder{b}{\sigma} \right) &= \frac{\varepsilon^2}{H^2} \pder{}{\sigma} \left( \kappa \pder{b}{\sigma} \right),
\end{align}
where $\varrho = N^2 H_0^2 / f^2 L^2$ is now the conventional Burger number.
Again assuming the scaling~$\varepsilon \ll 1$ and~$\mu \varrho \sim 1$, expanding interior variables in~$\varepsilon^2$, and matching orders as before, we arrive at the complete set of interior equations
\begin{align}
    -u^\eta\Izero &= -\pder{p\Izero}{\xi} + \sigma \pder{H}{x} b\Izero,\\
    u^\xi\Itwo &= \frac{1}{H^2} \pder{}{\sigma}\left(\nu \pder{u^\eta\Izero}{\sigma}\right),\label{eq:2d-eta-I2}\\
    \frac{1}{H} \pder{p\Izero}{\sigma} &= b\Izero,\\
    \pder{}{\xi}\left(H u^\xi\Itwo\right) + \pder{}{\sigma}\Big(H u^\sigma\Itwo\Big) &= 0,\\
    \mu \varrho \left( \pder{b\Izero}{t_2} + u^\xi\Itwo \pder{b\Izero}{\xi} + u^\sigma\Itwo \pder{b\Izero}{\sigma} \right) &= \frac{1}{H^2} \pder{}{\sigma} \left( \kappa \pder{b\Izero}{\sigma} \right).
\end{align}
We again find that the interior along-slope flow is of lower order than the interior cross-slope flow, and the interior buoyancy evolution is again slow.
In 2D, the interior slope-normal flow~$u^\sigma\Itwo$ comes in, contributing a second-order advective flux in the vertical, along with the cross-slope advection.
Formulated using the streamfunction~$\chi\Itwo$, this recovers the interior equations~\eqref{eq:2d-inversion-interior} and~\eqref{eq:2d-evolution-interior} derived above.
The $O(1)$ interior along-slope flow can again be obtained by integrating thermal wind in the vertical, with the bottom correction $u^\eta\Izero(-1)$ dropping out for $U^\xi = 0$.

The BL contribution can again be assessed after a re-scaling of the vertical coordinate such that~$\bar\sigma = \sigma/\varepsilon$.
We again find that the $O(1)$ BL flow, along with the interior along-slope flow~$u^\eta\Izero$ at the bottom, vanishes when~$U^\xi = 0$.
The BL flow is instead of~$O(\varepsilon)$, satisfying
\begin{align}
    -u^\eta\Bone &= -\pder{H}{x} b\Bone + \frac{1}{H^2} \pder{}{\bar\sigma}\left(\nu \pder{u^\xi\Bone}{\bar\sigma}\right),\\
    u^\xi\Bone &= \frac{1}{H^2} \pder{}{\bar\sigma}\left(\nu \pder{u^\eta\Bone}{\bar\sigma}\right),\\
    \mu \varrho u^\xi\Bone \pder{b\Izero}{\xi} &= \frac{1}{H^2} \pder{}{\bar\sigma} \left( \kappa \pder{b\Bone}{\bar\sigma} \right), \label{eq:2d-bl-buoyancy-nd}
\end{align}
with hydrostatic balance and continuity implying that~$p\Bone = 0$ and~$u^\sigma\Bone = 0$, respectively.
The BL is again characterized by a balance between cross-slope advection and down-gradient diffusion of buoyancy, with the BL buoyancy flux due to $b\Bone$ balancing the interior buoyancy flux due to $b\Izero$ at the bottom as before: $1 + \partial_\sigma b\Izero = -\partial_{\bar\sigma} b\Bone$ at~$\sigma = -1$.
The tendency term in \eqref{eq:2d-bl-buoyancy-nd} is again dropped because the interior evolution is slow, so the BL evolution must be slow as well.
Expressing $H u^\xi\Bone = \partial_{\bar\sigma} \chi\Btwo$, vertically integrating \eqref{eq:2d-bl-buoyancy-nd}, and enforcing $\chi\Itwo + \chi\Btwo = 0$ at $\sigma = -1$ yields an effective boundary condition on the interior.
The BL-interior exchange velocity~$u^\sigma\Itwo = -u^\sigma\Btwo$ at~$\sigma = -1$ may be obtained by vertically integrating
\begin{equation}
    \pder{}{\xi}\left( H u^\xi\Bone \right) + \pder{}{\bar\sigma}\Big( H u^\sigma\Btwo \Big) = 0.
\end{equation}
The leading-order equations obtained using this more rigorous approach again match the expressions derived heuristically above.
The asymptotic orders revealed by this approach are the same as in the 1D~case.

\section{Discussion}\label{s:discussion}

\cite{callies_dynamics_2018} studied the mixing-generated abyssal circulation in an idealized global basin using PG dynamics, but their model employed Rayleigh drag rather than a Fickian friction.
The models and theory presented here make use of a down-gradient turbulence closure of the momentum fluxes, allowing them to produce more realistic BLs and avoid unphysical interior momentum sinks.
Still, the results presented here provide some insight into the conclusions from this previous study. 
With Rayleigh drag, \citet{callies_dynamics_2018} found that the canonical 1D model was a reasonably accurate emulator for the full dynamics over slopes with a constant initial stratification.
This may have been somewhat of a coincidence, as in their case the steady state canonical transport~$\kappa_\infty \cot\theta$ was zero everywhere, adding a transport constraint to the canonical 1D model.
With Fickian friction, setting~$\kappa_\infty = 0$ does not immediately make the canonical 1D model equivalent to the transport-constrained 1D model because it still evolves diffusively and with nonzero transport, taking thousands of years to equilibrate \citepalias{peterson_rapid_2022}.
Rayleigh drag, in contrast, damps flow in the interior, allowing for fast adjustment (in a matter of years, not shown) to the~$U^\xi = 0$ steady state.
The combination of $\kappa_\infty = 0$ and Rayleigh drag thus conspired to let \citet{callies_dynamics_2018} get the right answer from the canonical model, but modifying either of these choices would have made the argument fall apart.

Furthermore, \citeauthor{callies_dynamics_2018}'s (\citeyear{callies_dynamics_2018}) application of BL theory was somewhat \textit{ad hoc}.
For slopes steep enough for the canonical BL theory to apply, the steady-state transport was exactly zero, meaning that all upslope transport was exactly balanced by downslope transport above.
The BL theory broke down at the base of the slopes, allowing the BLs to be fed by dense water from the south and the less dense downwelled water to return south, forming a basin-wide circulation that constituted an overturning.
The overturning transport could thus be estimated with an isobath integral of the upslope transport in BLs on the slopes.
As \citet{drake_abyssal_2020} pointed out, however, this approach is not successful if the interior stratification is far from constant and canonical BL theory does not apply.
The theory presented here supplies a globally valid expression for the BL transport that allows for variations in the interior stratification.
At this point, this expression is only a diagnostic tool, itself depending on the interior dynamics, but it unambiguously describes how the interior can exert control on the BL, and vice versa, ultimately generating a basin-wide circulation that involves both BL and interior pathways---and mass exchange between them.
This sharpens our view of the abyssal overturning, with no confusion about the roles of the BL and interior.

The framework presented here can also help understand the results from \citet{drake_abyssal_2020} regarding how water mass transformations are affected by changes in the interior stratification.
Using the same 3D PG model with Rayleigh friction as in \citet{callies_dynamics_2018}, they found that the degree of compensation between BL upwelling and interior downwelling is strongly dependent on vertical variations in the initial stratification.
With only the canonical 1D theory as a starting point, they were unable to explain the vertical extent and structure of water mass transformations.
The BL theory presented here would enable us to understand these physics more clearly, because it explicitly separates the BL and interior components of the flow.
This allows us to describe the abyssal circulation in terms of flows into and out of the BL, rather than simply bulk diapycnal motion throughout the water column.
In section~\ref{s:2d}, we demonstrated the power of this framework in describing abyssal spin up in 2D with exponential initial stratification.
Applied to 3D simulations such as those in \citet{drake_abyssal_2020}, this approach would undoubtedly shed light on what shapes the vertical structure of water mass transformations in the abyss.

Here, we have only presented results in 1D and 2D.
We leave the 3D case to a future paper, but preliminary work indicates that much of the theory developed here carries over, although there are some key differences.
In 3D, the interior dynamics satisfies geostrophic balance in both the~$\xi$ and~$\eta$ directions.
Because of this, the asymptotics in 3D are qualitatively different from those presented in section~\ref{s:asymptotics} of this paper: instead of evolving on a slower timescale, the leading-order 3D interior buoyancy field is advected by the geostrophic velocities, with diffusion only playing a role at higher order.
We anticipate that this qualitative difference between 2D and 3D may be crucial in explaining the full 3D abyssal circulation.
In 3D, it is also no longer possible to write the PG inversion in terms of a scalar streamfunction.
This makes the mathematics more complicated, but it is still possible to write down an expression for the 3D BL transport in terms of interior variables evaluated at the bottom.
As in 2D, the 3D BL mass and buoyancy transports feed back on the interior, now with gradients in the~$\eta$ direction shaping the flow field.
A future extension to 3D will allow us to explain the dynamics of abyssal circulations in more complicated and realistic geometries, including cases with variations in~$f$.\footnote{Variations in $f$ will allow for vortex stretching in the absence of friction: $\beta u^y = f \partial_z u^z$, where $\beta = \partial_y f$. 
In the $f$-plane solutions considered here, a non-zero interior vertical velocity only appears at second order (see section~\ref{s:asymptotics}).}

Our BL theory results are not only theoretically useful but could also lighten the computational demand of simulating the abyssal circulation.
The interior solution can be computed without the need to resolve the thin BL, allowing numerical models to have coarser grids and larger timesteps. 
This is crucial when studying the 3D system over long abyssal timescales of thousands or tens of thousands of years \citep[e.g.,][]{wunsch_how_2008,liu_transient_2009,jansen_transient_2018}.
This framework could even be used to analyze tracer transport without explicitly resolving the BL, allowing us to better understand carbon and heat storage \citep[e.g.,][]{sarmiento_new_1984} and Lagrangian pathways \citep[e.g.,][]{rousselet_coupling_2021} in the abyss.
If needed, the BL correction can be computed after the fact on a finer grid as was done for Figs.~\ref{fig:slope_profiles} and~\ref{fig:seamount_profiles}.

Although the results presented here are derived in the context of PG dynamics, they might also point the way towards a parameterization of the effects of BLs over a sloping seafloor in primitive-equation models.
Applying effective boundary conditions on the interior evolution, following the BL framework, should most easily be accomplished in models with terrain-following coordinates
But a translation to $z$-coordinates also appears feasible, which would alleviate not only the need to resolve thin boundary layers in the vertical but also the need to capture BL flow across the artificial steps in the topography in such models.
An extension of the BL theory to 3D is needed, however, to produce expressions directly useful for such a parameterization effort.

The circulation in the examples presented in this paper depend on the particular, simple closure of turbulent momentum and buoyancy fluxes employed in all of them.
Although Fickian friction is much more physical than Rayleigh drag, our use of it with a simple profile for~$\nu$ still glosses over the true complexity of turbulence in the abyss.
Without a more faithful representation of the internal-wave field and baroclinic eddies in abyssal mixing layers, we cannot claim to be accurately simulating the dynamics of the real ocean.
The BL framework, however, is robust to the choice of turbulence parameterization---as long as the vertical scale of the turbulent mixing in the interior is larger than the thickness of the BL, our approach should require minimal modification.
The results presented here are in terms of a particular choice of parameterization, but the general themes describing how the BL and interior communicate will carry over to more complex closures.
This flexibility makes BL theory an attractive tool for understanding the mixing-generated abyss over a hierarchy of complexities.
 
\section{Conclusions}\label{s:conclusions}

Motivated by observations of bottom-enhanced mixing, recent work on the abyssal circulation has focused on the role of thin bottom BLs \citep{ferrari_turning_2016,de_lavergne_consumption_2016,mcdougall_abyssal_2017,holmes_ridges_2018,callies_dynamics_2018,drake_abyssal_2020}.
Until now, the coupling between these BLs and the interior circulation remained opaque, with most of our understanding coming from somewhat heuristic arguments using 1D theory.
The framework presented in this work uses BL theory to paint a clear picture of the interior--BL interaction of the mixing-generated abyssal circulation. 
By explicitly defining BL and interior contributions to the flow, we obtain expressions for the BL transport in 1D and 2D that are bounded for all bottom slopes, solving the old 1D conundrum of the steady total transport~$\kappa_\infty \cot\theta$ being set by the far-field mixing and diverging for small slopes.
In the revised theory, the BL transport is set by local flow parameters and interior variables evaluated at the bottom, with the total transport allowed to evolve according to the global context.
The interior dynamics are themselves modified by this BL transport, which advects dense water up-slope and thus modifies the interior bottom boundary condition.
This two-way coupling provides a complex yet transparent story of how BLs influence the abyssal circulation, and this framework makes previously unwieldy problems, such as determining the response to vertically varying initial stratification, comparatively simple.
With these promising results, we anticipate that BL theory will play a crucial role in the development of a more complete understanding of the abyssal circulation in the real ocean. 

\datastatement
The numerical models for all the simulations presented here are hosted at \url{https://github.com/hgpeterson/nuPGCM}.

\acknowledgments
This material is based upon work supported by the National Science Foundation under Grant No.~OCE-2149080.
We thank David Marshall and an anonymous reviewer for their helpful feedback on our original manuscript.

\clearpage
\appendix[A]
\appendixtitle{BL Theory when $U^\xi \neq 0$}

\noindent For completeness, we here show how the BL theory derivations in sections~\ref{s:1d} and~\ref{s:2d} are slightly altered when the transport~$U^\xi$ is non-zero.
In both 1D and 2D, the interior inversion is modified to include the added transport term.
The BL accounts for~$1 / (1 + \mu \varrho)$ of the total transport, leading to a modified interior bottom boundary condition compared to before.
The 2D case is special in that the total transport is itself a function of the flow and geometry of the domain (see appendix~B of \citetalias{peterson_rapid_2022}), allowing us to derive an explicit equation for~$U^\xi$ in that case.

\subsection{One-dimensional theory}

The 1D interior inversion for general~$U^\xi$ is
\begin{equation}\label{eq:1d-inversion-interior-Uneq0}
    \frac{f^2}{\nu} ( \chi\I - U^\xi ) = -\pder{b'\I}{\zeta} \tan\theta.
\end{equation}
This does not affect~$\partial_\zeta \chi\I$, leaving the interior evolution equation~\eqref{eq:1d-evolution-interior} unchanged.
This new interior balance results in a modified bottom boundary condition compared with equation~\eqref{eq:1d-interior-bc}:
\begin{equation}
    \kappa \left[ N^2 + (1 + \mu \varrho) \pder{b'\I}{\zeta} \right] =  U^\xi N^2 \tan\theta \quad \mathrm{at} \quad \zeta = 0.
\end{equation}
The added flux on the right-hand side represents the integrated buoyancy supplied to the column by the net transport~$U^\xi$.
The BL transport (cf.~\ref{eq:1d-bl-transport}) now takes the form
\begin{equation}\label{eq:1d-bl-transport-Uneq0}
    \chi\I = \frac{U^\xi}{1 + \mu \varrho} + \kappa \cot\theta \frac{\mu \varrho}{1 + \mu \varrho} \quad \mathrm{at} \quad \zeta = 0,
\end{equation}
supplying a fraction of the total transport.
For~$\mu \varrho \ll 1$, the BL absorbs the majority of the added transport.
Note that the bottom boundary condition may be written as~$\kappa(N^2 + \partial_\zeta b'\I) = \chi\I N^2 \tan\theta$ regardless of whether~$U^\xi$ is nonzero.
The BL correction~$\chi\B$ remains the same as in~\eqref{eq:1d-bl-sol}, with~$\chi\I$ at~$\zeta = 0$ now coming from~\eqref{eq:1d-bl-transport-Uneq0}.

The asymptotic order of $U^\xi$ must match that of $\chi\I$, so it must be restricted to be $O(\varepsilon^2)$. It is then simple to incorporate $U^\xi \neq 0$ into the theory presented in section~\ref{s:asymptotics}.

\subsection{Two-dimensional theory}

In 2D, the general interior inversion is
\begin{equation}
    \frac{f^2}{\nu} ( \chi\I - U^\xi ) = \pder{b\I}{\xi} - \frac{\sigma}{H} \pder{H}{x} \pder{b\I}{\sigma},
\end{equation}
and again the BL absorbs a fraction of the added transport so that~\eqref{eq:2d-bl-transport} becomes
\begin{equation}\label{eq:2d-bl-transport-Uneq0}
    \chi\I = \frac{U^\xi}{1 + \mu \varrho} + \frac{\kappa}{\partial_x H} \frac{\mu \varrho}{1 + \mu \varrho} \quad \mathrm{at} \quad \sigma = -1,
\end{equation}
where~$\varrho = -\partial_x H \partial_\xi b\I / f^2$ at~$\sigma = -1$.
For symmetric topography, $U^\xi = 0$, but this is not the case in general.
We can infer~$U^\xi$ for asymmetric geometries with knowledge of the interior buoyancy distribution.
Evaluating~\eqref{eq:2d-ueta-interior} at~$\sigma = 0$ and taking the mean in~$\xi$, denoted by~$\langle \cdot \rangle$, we have
\begin{equation}\label{eq:along-slope-integral}
    \left\langle u^\eta\I(0) \right\rangle = 0 = -\left\langle \frac{f}{q \nu} \chi\I \right\rangle + \left\langle \frac{H}{f} \int_{-1}^0 \pder{b\I}{x}(\sigma) \; d\sigma \right\rangle,
\end{equation}
where, crucially, the BL transport from equation~\eqref{eq:2d-bl-transport-Uneq0} now depends on~$U^\xi$ .
We have assumed that the domain is tall enough such that gradients in buoyancy at~$\sigma = 0$ are small and therefore~$\langle u^\eta\I(0) \rangle = 0$.
Solving for~$U^\xi$ yields
\begin{equation}
    U^\xi = \frac{ 
        \left\langle H \int_{-1}^0 \pder{b\I}{x}(\sigma) \; d\sigma \right\rangle 
        + 
        \left\langle \frac{1}{q} \frac{\partial_\xi b\I}{1 + \mu \varrho} \right\rangle
        }
        {
        \left\langle \frac{f^2}{q \nu} \frac{1}{1 + \mu \varrho} \right\rangle
        },
\end{equation}
where all variables are evaluated at~$\sigma = -1$ unless otherwise noted.
Simulations of an asymmetric ridge, similar to that in appendix~B of \citetalias{peterson_rapid_2022}, confirm the accuracy of this formula (not shown).

Again, we restrict ourselves to cases where the non-dimensional~$U^\xi$ is~$O(\varepsilon^2)$, the same order as~$\chi\I$.
This is true when the second term on the right in equation~\eqref{eq:along-slope-integral} of lower order than the first.
This is always the case after a fast initial adjustment.

\clearpage
\appendix[B]
\appendixtitle{Axisymmetric Coordinates}

\noindent For simulations of an idealized seamount, we transform to axisymmetric coordinates, assuming rotational symmetry.
The depth~$H$ is then a function of the radial distance~$r$ and invariant under rotation about the origin by some angle~$\phi$, leading to effectively 2D flow.
Defining~$\rho = r$ and~$\sigma = z/H$, we have
\begin{align}
    -\rho f u^\phi &= -\pder{p}{\rho} + \sigma\pder{H}{r} b + \frac{1}{H^2}\pder{}{\sigma}\left(\nu\pder{u^\rho}{\sigma}\right),\\
    \rho f u^\rho &= \frac{\rho^2}{H^2}\pder{}{\sigma}\left(\nu\pder{u^\phi}{\sigma}\right),\\
    \pder{p}{\sigma} &= b H,\\
    \pder{}{\rho}\left(\rho H u^\rho\right) + \pder{}{\sigma}\left(\rho H u^\sigma\right) &= 0,\\
    \pder{b}{t} + u^\rho\pder{b}{\rho} + u^\sigma\pder{b}{\sigma} &= \frac{1}{H^2}\pder{}{\sigma}\left(\kappa\pder{b}{\sigma}\right).
\end{align}
The streamfunction inversion takes the same form as in Cartesian coordinates,
\begin{equation}
    \frac{1}{H^4}\pder{^2}{\sigma^2}\left(\nu\pder{^2\chi}{\sigma^2}\right) + \frac{f^2}{\nu}(\chi - U) = \pder{b}{\rho} - \frac{\sigma}{H}\pder{H}{r}\pder{b}{\sigma},
\end{equation}
with a slight difference in the streamfunction definition due to the new form of the divergence operator:
\begin{equation}
    u^\rho = \frac{1}{H}\pder{\chi}{\sigma} \quad \text{and} \quad u^\sigma = -\frac{1}{\rho H}\pder{(\rho\chi)}{\rho}.
\end{equation}

\clearpage
\bibliographystyle{ametsocV6}
\bibliography{biblio}

\end{document}